\documentclass[12pt]{JHEP} 

\usepackage{epsfig}

\title{Signal of neutrinoless double beta decay, neutrino spectrum and
oscillation scenarios}

\author{Francesco Vissani\\
	Deutsches Elektronen-Synchrotron, DESY\\
	Notkestra\ss{}e 85, D-22603 Hamburg, Germany, and\\
	International Centre for Theoretical Physics, ICTP\\
	Strada Costiera 11, 34100 Trieste, Italy\\
	E-mail: \email{vissani@ictp.trieste.it}}


\abstract{The lower  and upper bounds on the  neutrinoless double beta
($0\nu  2\beta$)  decay  rate   are  obtained,  as  functions  of  the
parameters of neutrino oscillations and of the lightest neutrino mass.
The  constraints on  these parameters  from the  search for  the $0\nu
2\beta$ transition,  as well as  from the interpretation of  solar and
atmospheric   neutrino  data   in  terms   of  oscillations,   can  be
conveniently   represented   in    one   unitarity   triangle.    This
representation helps to clarify the  cases when the $0\nu 2\beta$ rate
is small;  the crucial dependence  on the scenarios assumed  for solar
neutrino oscillations and on  the neutrino spectrum is emphasized.  We
consider  hierarchical  and  non-hierarchical  neutrino  spectra,  and
discuss their interest in view of future searches of the $0\nu 2\beta$
decay.}  

\begin{document}

\section{Informations on neutrino parameters} 

\subsection{Massive neutrinos and $0\nu 2\beta$ decay}

Atmospheric neutrino data can be interpreted in terms of a dominant
$\nu_\mu-\nu_\tau$ oscillation channel, although a sub-dominant
channel $\nu_\mu-\nu_{\rm e}$ is not excluded~\cite{subdmix}.  The
latter may be due to a $\nu_e$ component of the heaviest (lightest)
neutrino state $\nu_3$ ($\nu_1$) for spectra with ``normal''
(``inverted'') hierarchy--our definition of ``hierarchy'' is discussed
in section~\ref{secdef}.  Several possibilities are open for the
interpretation of the solar neutrino data, depending on the
frequencies of oscillation and mixings.

Hence, the indications for massive neutrinos are strong.  However,
there is quite a limited knowledge on the {\em neutrino mass spectrum
itself}, and particularly on the lightest neutrino mass.  The search
for $0\nu 2\beta$ decay can shed light on this important issue.  The
bound of 0.2 eV obtained~\cite{2b0nuexp} on the parameter
\begin{equation}
{\cal M}_{{\rm ee}}=| \sum_i U_{{\rm e}i}^2\ m_i |
\label{eq1}
\end{equation}
is sensibly smaller than the mass scales probed by present studies of
$\beta$-decay, or those inferred in cosmology~\cite{PDG}.  In
eq.~(\ref{eq1}), the non-negative quantities $m_i,$ $i=1,2,3...N$ are
the neutrino masses ($m_{i+1}\ge m_i$); the complex quantities
$U_{\ell i},$ $\ell={\rm e},\mu,\tau...,$ are the elements of the
mixing matrix, which relates the flavor eigenstates to the mass
eigenstates: $\nu_\ell(x)=\sum_i U_{\ell i}\, \nu_i(x).$ Hence, ${\cal
M}_{{\rm ee}}$ can be thought of as (the absolute value of) the
ee$-$entry of the neutrino mass matrix.  Let us recall that, beside
the $(N-1)(N-2)/2$ phases relevant to neutrino oscillations, there are
still $N-1$ physical phases in the lepton sector that have no analogy
in the quark sector, and arise from the Majorana structure of the
neutrino mass matrix.  Notice that both the amplitudes {\em and} the
phases of the elements of the mixing matrix $U_{{\rm e}i}$ are
relevant in determining the size of ${\cal M}_{{\rm ee}}.$

\subsection{Extremal values of ${\cal M}_{{\rm ee}}$ for $0\nu2\beta$ decay}

We obtain in this section the extremal values of ${\cal M}_{\rm ee}$
under arbitrary variations of the phases, keeping fixed the neutrino
masses $m_i$ and the ``mixing elements''\footnote{In the following, we
will always refer with the term ``mixing elements'' to the absolute
value of the elements of the mixing matrix.}  $|U_{{\rm e}i}^2|.$ The
maximum value of ${\cal M}_{\rm ee}$ is simply:
\begin{equation}
{\cal M}_{\rm ee}^{max} = \sum_i |U_{{\rm e}i}^2 |\ m_i . 
\label{eq2}
\end{equation}
The minimum value can  be written as:
\begin{equation}
 {\cal M}_{\rm ee}^{min} = 
{\rm max}\{\ 2\ |U_{{\rm e}i}^2|\ m_i-
{\cal M}_{\rm ee}^{max}
 ,\ \  0 \ \}.
\label{eq3}
\end{equation}
To demonstrate this formula, let us consider the absolute value of the
sum of three complex numbers: $r=|z_1+z_2+z_3|.$ We want to minimize
$r$ by keeping fixed $|z_i|,$ namely, by varying the phases.  Let us
define the quantities $r_{1,2,3}$ and $q_{1,2,3}$ as: $r_1=|z_1| -
|z_2| - |z_3|,$ $q_1=|z_1|-|z_2+z_3|,$ and similar eqs., but permuting
the indices for $r_{2,3}$ and $q_{2,3}.$ Notice that {\em at most} one
of the $r_i$`s is positive.  Assuming that $r_1>0,$ it is simple to
show that $r^{min}=r_1;$ in fact, using twice the Schwartz inequality,
we get $r\ge | q_1 |=q_1 \ge r_1.$ Similar considerations if $r_2>0,$
or $r_3>0.$ The last case has $r_i\le 0$ for $i=1,2,3.$ If one of the
$r_i$`s is zero, then $r^{min}=0,$ hence we need to consider the case
when $r_i<0$ for all $i$`s. In this case, the quantity $q_1$ goes from
negative, when the phases of $z_2$ and $z_3$ are equal, to positive,
when these phases are opposite. By continuity, a phase choice exists
such that $q_1=0.$ Since by proper choice of the phase of $z_1$ we can
get $r=|q_1|$ we conclude that, again, $r^{min}=0.$ In conclusion, the
general case is covered by the formula: $r^{min}=\mbox{max}\{r_i,\ 0
\}.$ This is equivalent to eq.~(\ref{eq3}), after noticing that $r_i=2
|z_i|-\sum_{i=1}^3 |z_i|.$ The generalization of these results to $N$
neutrinos is quite simple: Just limit the sum in eq.~(\ref{eq2}) to
$N=3.$ However, we will be concerned only with the case of three
neutrinos in the rest of the work.

The previous two equations give the extremal values of ${\cal M}_{\rm
ee},$ once the neutrino spectrum {\em and} the mixing elements are
known.  Such extremal values are important, being independent of the
complex phases.  The information we get from the experimental upper
bound is ${\cal M}_{\rm ee}^{bound} \ge {\cal M}_{\rm ee}^{min};$ the
informations we could get from a positive signal, instead, is ${\cal
M}_{\rm ee}^{signal} \in [{\cal M}_{\rm ee}^{min},\ {\cal M}_{\rm
ee}^{max}].$ In the following it will be shown how to use and
represent ${\cal M}_{\rm ee}^{min}$ and ${\cal M}_{\rm ee}^{max},$ and
what we can learn on them assuming specific neutrino spectra, and
scenarios of neutrino oscillations.

\section{Representation of ${\cal M}_{\rm ee}^{min}$ and 
${\cal M}_{\rm ee}^{max}$}

We introduce and discuss in this section a graphical representation of
the values of ${\cal M}_{\rm ee}^{min}$ and ${\cal M}_{\rm ee}^{max}.$
For this purpose we will make reference to fig.~\ref{f:1}, where the
representation of ${\cal M}_{\rm ee}^{min}$ is displayed, for an
illustrative choice of the neutrino spectrum: $m_3=2\ m_2$ and $m_2=2\
m_1.$ In order to fix the ideas, we point out from the beginning the
two essential features of fig.~\ref{f:1}: (1) the value of $ {\cal M}_{\rm
ee}$ at the vertices, namely the masses of the neutrinos $m_i$; (2)
the position of the inner triangle (also determined by the masses of
the neutrinos).

\FIGURE{\epsfig{file=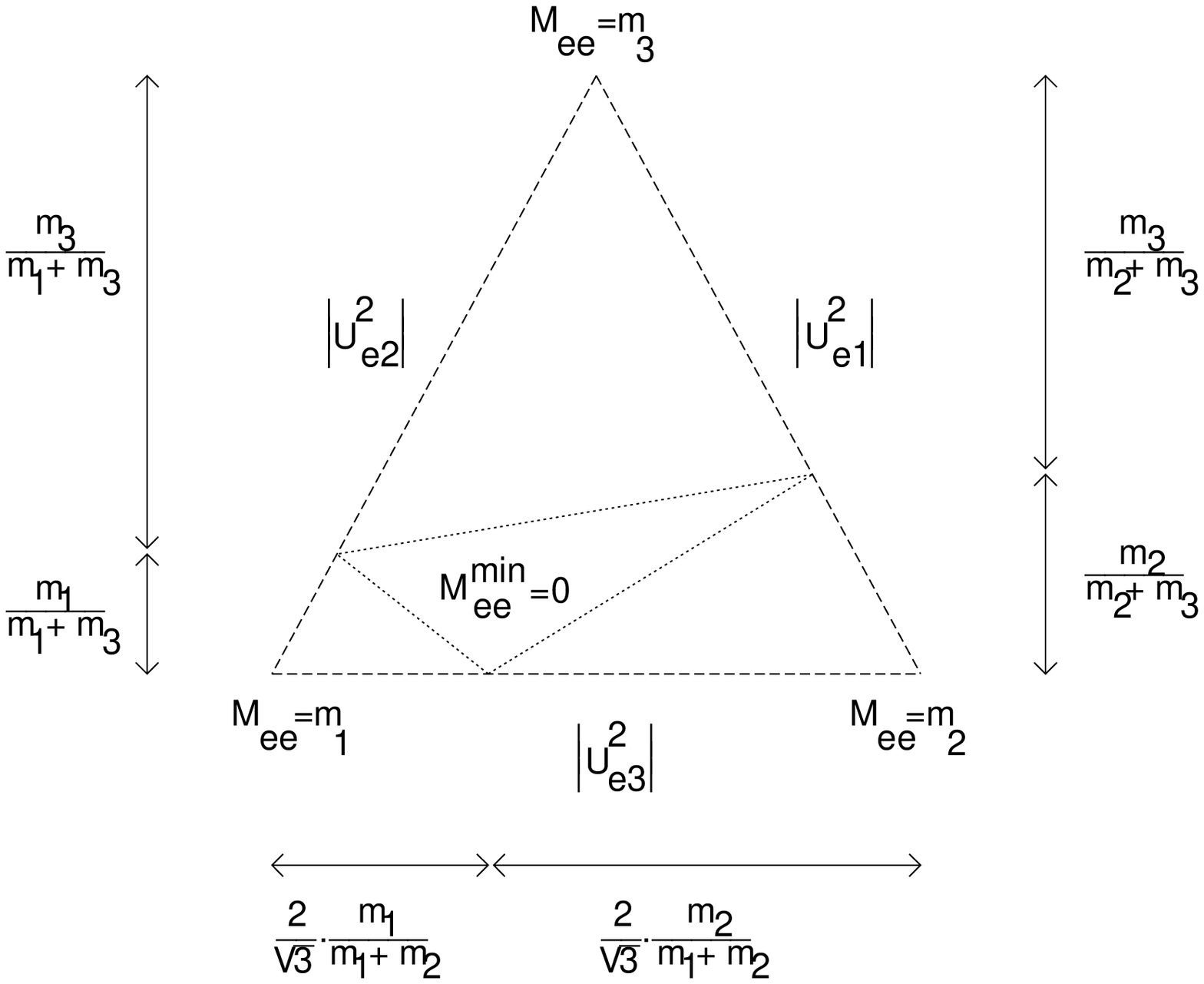,angle=270,width=29em}
\caption{Representation of the minimum value of {${\cal M}_{\rm ee},$}
eq.~({\ref{eq3}}), in the unitarity triangle.  For a given internal
point, the distance from the side labelled by {$|U_{{\rm e}i}^2|$}
represents the size of the corresponding mixing element.  The inner
triangle encloses the region where {${\cal M}_{\rm ee}^{min}=0.$} The
position of the vertices of the inner triangle relative to the
vertices of the unitarity triangle (corresponding to
eq.~({\ref{eq4}})) is indicated by the arrows with
labels.\label{f:1}}}

Let us begin by recalling some basic facts. The three mixing elements
$|U_{{\rm e}i}^2|$ are constrained by the unitarity condition $ \sum_i
|U_{{\rm e}i}^2|=1.$ This condition can be represented by using the
inner region of one equilateral triangle with unit height, where the
distance from the $i^{th}$ side represents the value of $|U_{{\rm
e}i}^2|,$ see fig.~\ref{f:1} (this triangle was first used in
\cite{1sttri}, to analyze solar neutrino oscillations).  To exemplify
the use of the triangle, let us consider two special cases: (a) When
$\nu_{\rm e}$ is an equal admixture of the three mass eigenstates, we
have $|U_{{\rm e}i}^2|=1/3.$ This point is represented by the
barycentre of the equilateral triangle of fig.~\ref{f:1}.  (b) When
$\nu_{\rm e}$ coincides with the mass eigenstate $\nu_1,$ we have
$|U_{{\rm e}1}^2|=1,$ and the other two mixing elements are zero.
This point is represented by the $1^{st}$-vertex (by definition, the
$1^{st}$ vertex is opposite to the $1^{st}$ side, denoted with the
label $|U_{{\rm e}1}^2|$ in fig.~\ref{f:1}, {\em etc.}).

{}From eq.~(\ref{eq3}), $ {\cal M}_{\rm ee}^{min}$ is {\em zero} in
the inner triangular region represented in fig.~\ref{f:1}.  The vertices of
this inner triangle are given by:
\begin{equation}
|U_{{\rm e}1}^2|/|U_{{\rm e}2}^2|=m_2/m_1 \mbox{  when  } |U_{{\rm e}3}^2|=0 ,
\label{eq4}
\end{equation}
and by the two additional equations obtained by the replacement
$3\leftrightarrow 1,$ and $3\leftrightarrow 2.$ The condition
$|U_{{\rm e}3}^2|=0$ in eq.~(\ref{eq4}) tells us that we are on the
$3^{rd}$ (lower) side of the unitarity triangle of fig.~\ref{f:1}.

At the $i^{th}$ vertex of the unitarity triangle ${\cal M}_{\rm
ee}^{min}={\cal M}_{\rm ee}=m_i,$ as is clear from eq.~(\ref{eq3}),
and as illustrated in fig.~\ref{f:1}.  The value of ${\cal M}_{\rm ee}^{min}$
decreases linearly when moving from one vertex toward the inner
triangle.  In fact, $ {\cal M}_{\rm ee}^{min}$ is non-zero only close
to the vertices of the unitarity triangle (assuming $m_1>0$).  This
concludes the illustration of fig.~\ref{f:1}.

The unitarity triangle can also be used to represent the maximum
possible value ${\cal M}_{\rm ee}.$ Quite simply, ${\cal M}_{\rm
ee}^{max}$ is the function of the mixing elements $|U_{{\rm e}i}^2|$
that interpolates linearly among the values ${\cal M}_{\rm ee}=m_i$
taken at the vertices of the unitarity triangle, as clear from
eq.~(\ref{eq2}). However, since ${\cal M}_{\rm ee}^{max}$ is just the
sum of positive contributions (eq.~(\ref{eq2})), the analysis of
${\cal M}_{\rm ee}^{max}$ is nearly trivial.

\section{Phenomenology of oscillations and $0\nu 2\beta$}\label{secdef}

We discuss now the $0\nu 2\beta$ signal assuming some specific
spectra, and scenarios of oscillation, using the graphical
representation introduced above.  We take advantage of the indications
from atmospheric and solar neutrinos, that can be accounted in terms
of two different frequencies of neutrino oscillations, related to the
mass differences squared $\Delta m_{atm}^2$ and $\Delta m_{\odot}^2$
($\Delta m_{atm}^2\gg \Delta m_{\odot}^2$).  We consider the following
three cases:
\begin{itemize}
\item{} Case [${\cal N}$]: ``normal'' hierarchy, 
                    $m_1\ll (\Delta m_{atm}^2)^{1/2}$;
\item{} Case [${\cal I}$]: ``inverted'' hierarchy,
                    $m_1\ll (\Delta m_{atm}^2)^{1/2}$; 
\item{} Case [${\cal D}$]: ``normal'' and ``inverted'' hierarchies, 
                    $m_1\gg (\Delta m_{atm}^2)^{1/2}$; 
\end{itemize}
from these cases, it will be easy to understand also the
``intermediate'' situations when $m_1\sim (\Delta m_{atm}^2)^{1/2}.$
With the term ``hierarchy'' (either ``normal'' or ``inverted'') we
refer to {\em the mass differences squared} (see eqs.~ (\ref{defnh})
and (\ref{defih}) below)\footnote{In order to simplify the connection
with the phenomenology, we use a definition of ``hierarchy'' that is
relevant to neutrino oscillations, which involves {\em just} the mass
differences squared.  Notice that sometimes in the literature,
``hierarchy'' is used in reference to the neutrino spectrum itself.}.
We assume that the electronic admixture in atmospheric neutrinos is
sub-dominant~\cite{subdmix}, and use for the mass splittings $\Delta
m_{atm}^2$ and $ \Delta m_{\odot}^2$ the values suggested by the
phenomenology.  For solar neutrino solutions we use the terminology of
\cite{bks}, that we will recall in the following.  A similar study has
been performed in reference~\cite{fmn}, with the goal to extract
informations on the mixing angles, knowing ${\cal M}_{{\rm ee}}$ and
the neutrino spectrum. For other recent works oriented toward the
phenomenology, see~\cite{more}.

\subsection{Case \protect{[${\cal N}$]}:
``normal'' hierarchy, \protect{$m_1\ll (\Delta m_{atm}^2)^{1/2}$}}
\label{sec-n} What is the expected value of ${\cal M}_{{\rm ee}}$ for
a neutrino spectrum with ``normal'' hierarchy:
\begin{equation}
m_3^2-m_2^2=\Delta m^2_{atm}\gg m_2^2-m_1^2=\Delta m^2_{\odot},
\label{defnh}
\end{equation}
assuming, to begin with, that $m_1$ is negligible?  For the values of
$\Delta m^2_{\odot}$ suggested by the MSW~\cite{MSW} small mixing
angle solution of the solar neutrino problem (SMA) or vacuum
oscillation (VO), the only important contribution to $0\nu 2\beta$
decay rate comes from the heaviest eigenstate: ${\cal M}_{{\rm
ee}}\approx |U_{{\rm e}3}^2| m_3.$ It is {\em possible} to have a
comparable contribution from the second eigenstate assuming MSW
solutions of the solar neutrino problem with large mixing angle (LMA)
$\delta {\cal M}_{{\rm ee}} |_{\odot} = |U_{\rm e2}^2|\ m_2 \approx
4\times 10^{-3}$ eV (using $\Delta m^2_{\odot}\approx 10^{-4}$ eV$^2$
and $|U_{\rm e2}^2|\approx 0.4$).  This is of the same size of the
contribution from the heaviest eigenstate, $\delta {\cal M}_{{\rm ee}}
|_{atm}=|U_{\rm e3}^2|\ m_3,$ if $|U_{\rm e3}^2|\approx 0.1 $ and
$\Delta m^2_{atm}\approx 2\times 10^{-3}$ eV$^2.$ We conclude that, if
future experiments searching for the $0\nu 2\beta$ transition will
prove that
\begin{equation}
{\cal M}_{{\rm ee}} > 10^{-2} \mbox{ eV} ,
\label{normsig}
\end{equation}
the hypothesis of a spectrum with ``normal'' hierarchy and very small
$m_1$ will be disfavoured~\cite{[b1]}\footnote{Alternatively, one
should postulate a different origin of the $0\nu 2\beta$ decay.}.

\FIGURE{%
\epsfig{file=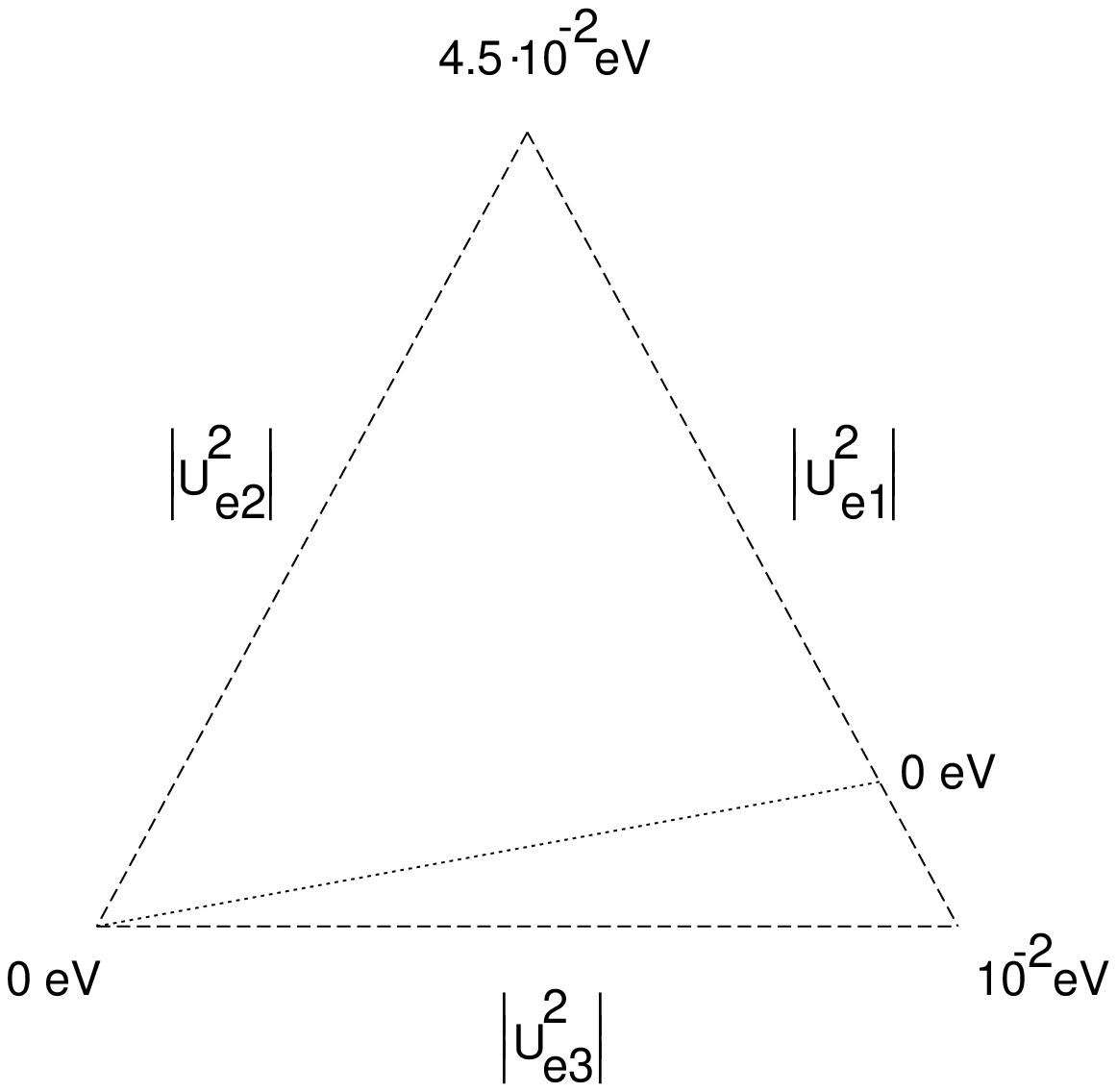,angle=270,width=18em}%
\epsfig{file=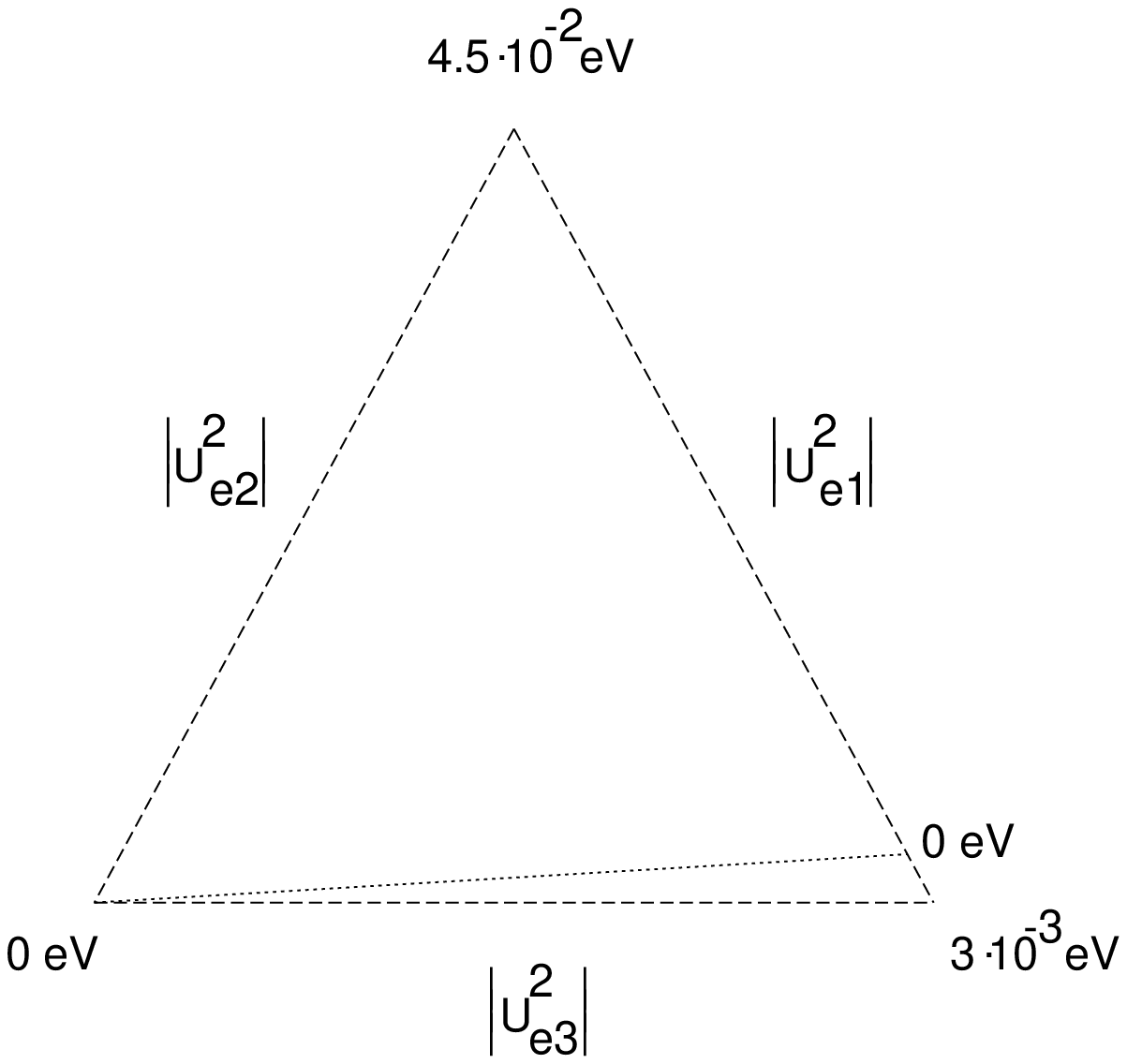, angle=270,width=18em}%
\caption{Same as fig.~\ref{f:1}, for spectra with {$m_1=0$} and
``normal'' hierarchy.  For ``inverted'' hierarchy, the value at the
{$2^{nd}$} vertex increases from the value {$10^{-2}$} in the first
plot ({$3\times 10^{-3}$} in the second plot) up to {$\approx
4.5\times 10^{-2}$} eV, and the internal line approaches closely the
bisector {$|U_{{\rm e}2}^2| = |U_{{\rm e}3}^2|.$}\label{f:2}}}

The function ${\cal M}_{{\rm ee}}^{min}$ is represented in
fig.~\ref{f:2} for two different values of $\Delta m^2_{\odot}:$
$10^{-4}$ eV$^2$ in the $1^{st}$ plot, and $10^{-5}$ eV$^2$ in the
$2^{nd}$ (we assumed $\Delta m^2_{atm}=2\times 10^{-3}$ eV$^2$).
Notice that assuming $m_1=0$ the inner triangle of fig.~\ref{f:1}
degenerates into a line (for much smaller values of $\Delta
m^2_{\odot},$ say for VO, the line practically coincides with the side
$U_{{\rm e}3}=0$).  Recalling that the inner triangle corresponds to
the region where ${\cal M}_{{\rm ee}}^{min}=0,$ we appreciate from
fig.~\ref{f:2} the crucial dependence on the parameter $|U_{{\rm
e}3}^2|$ of the $0\nu2\beta$ transition rate.

\pagebreak[3]

Let us increase now the size of $m_1,$ keeping $m_1 \ll (\Delta
m^2_{atm})^{1/2}\approx m_3.$ The (degenerate) inner triangle in
fig.~\ref{f:2} becomes an {\em obtuse} isosceles triangle when
$m_1\approx m_2 \raisebox{-.4ex}{\rlap{$\sim$}} \raisebox{.4ex}{$>$}
(\Delta m^2_{\odot})^{1/2};$ the base being parallel to the $3^{rd}$
side, where $U_{{\rm e}3}=0.$ A complete suppression of the $0\nu
2\beta$ transition can take place for those solutions of the solar
neutrino problem that fall in this inner triangle, and for this
reason, the most important conclusion is unchanged: The size of
$|U_{{\rm e}3}^2|$ is very important in determining whether the case
${\cal M}_{{\rm ee}}^{min}=0$ is possible or not.  More precisely,
this mixing element has to be compared with the height of the
triangle, $\sim m_1/m_3$ (see fig.~\ref{f:1}).  Incidentally, we
notice the simple formula
\begin{equation}
{\cal M}_{{\rm ee}}\approx |m_1 + |U_{{\rm e}3}^2|\  m_3\ e^{i \varphi}|
\ \ \ \mbox{ where } \varphi={\rm arg}[\, U_{{\rm e}3}^2/U_{{\rm e}1}^2\, ]
\label{smanorm}
\end{equation}
valid for the SMA case, which illustrates that ${\cal M}_{{\rm
ee}}\approx 0$ is possible when $m_1/m_3 \approx |U_{{\rm e}3}^2|$
($m_3\approx (\Delta m^2_{atm})^{1/2}$ in present hypotheses) and the
phases of $U_{{\rm e}3}^2$ and $U_{{\rm e}1}^2$ are opposite.

\subsection{Case {[${\cal I}$]}: ``inverted'' hierarchy, 
{$m_1\ll (\Delta m_{atm}^2)^{1/2}$}}
\label{sec-i} 

Let us assume a spectrum with ``inverted'' hierarchy, namely
\begin{equation}
m_3^2-m_2^2=\Delta m^2_{\odot} \ll m_2^2-m_1^2=\Delta m^2_{atm},
\label{defih}
\end{equation}
and suppose, to begin with, that $m_1$ is negligible.  In this case,
since the sub-dominant mixing element is $|U_{{\rm e}1}^2|,$ we can
obtain large maximum values~\cite{[b2]}:
\begin{equation}
{\cal M}_{{\rm ee}}^{max} 
\approx (\Delta m^2_{atm})^{1/2} 
=(3 \mbox{ to } 9) \times 10^{-2}  \mbox{ eV.}
\label{invsig}
\end{equation}
This could be close to the present bound~\cite{2b0nuexp}, if also the
nuclear matrix elements take the highest values allowed by present
uncertainties, $\sim 2-3$~\cite{simk}.

In these hypotheses, ${\cal M}_{{\rm ee}}^{min}$ can be (close to)
zero only if $|U_{{\rm e}2}^2|$ is very close to $|U_{{\rm e}3}^2|;$
the contribution from $|U_{{\rm e}1}^2|$ being irrelevant.  In a
graphical representation like in fig.~\ref{f:2}, this corresponds to
the fact that the inner triangle almost coincides with the bisector
$|U_{{\rm e}2}^2| = |U_{{\rm e}3}^2|$ (the ``small'' mixing element
$|U_{{\rm e}1}^2|$ is represented by the distance from the
$1^{st}$--right--side).

Let us increase the size of $m_1,$ keeping $m_1 \ll (\Delta
m^2_{atm})^{1/2}\approx m_3.$ The inner triangle is, in this
assumption, {\em acute} isosceles, the base being parallel to the side
$U_{{\rm e}1}=0,$ and with length $\sim m_1/m_3\times 2/\sqrt{3}.$
Hence, only those solutions of the solar neutrino problem which have
almost maximal mixing angles (VO, averaged oscillations and perhaps
LMA) fall in the region where the $0\nu 2\beta$ transition rate may be
strongly suppressed.  In the case of SMA, since $|U_{\rm e3}^2|$ is
small by assumption (and $|U_{\rm e1}^2|$ is not large) we have
simply:
\begin{equation}
{\cal M}_{{\rm ee}}\approx m_2\approx 
(m_1^2 + \Delta m^2_{atm})^{1/2}.
\label{smainverted}
\end{equation}
Hence, ${\cal M}_{{\rm ee}}\approx 0$ is impossible if the SMA
solution is correct.  Quite generally, in the case of ``inverted''
hierarchy, it is less likely that ${\cal M}_{{\rm ee}}^{min}$ is zero.

\subsection{Case {[${\cal D}$]}: ``nearly degenerate'' spectrum,
{$m_1\gg (\Delta m_{atm}^2)^{1/2}$}}\label{sec-d}

Largest values of ${\cal M}_{{\rm ee}}$ (up to the experimental bound)
can be taken for a ``nearly degenerate'' neutrino spectrum
\cite{[1],[2]}.  The maximum value is simply ${\cal M}_{{\rm
ee}}^{max}=m_1+{\cal O}(\Delta m^2/m_1),$ $m_1$ playing the role of
mass spectrum offset.

The corresponding minimum value, ${\cal M}_{{\rm ee}}^{min}/m_1={\rm
max}\{ 2 |U_{{\rm e}i}^2|-1,\ 0 \}$ is represented in fig.~\ref{f:3}
assuming ``normal'' hierarchy of the mass differences
(eq.~(\ref{defnh})); ${\cal O}(\Delta m^2/m_1^2)$ terms have been
neglected.  {}From this figure it is visible that, to interpret
properly the results of $0\nu 2\beta$ decay studies (and possibly, to
exclude the inner region in the $1^{st}$ plot, the one where ${\cal
M}_{{\rm ee}}\ll m_1$ is {\em possible}) we need precise information
on the mixing elements.  This requires distinguishing among
oscillation scenarios.  The plots also illustrate the importance to
quantify the size of $|U_{\rm e3}^2|$ \cite{[1],[2]},~\cite{subdmix}.
Similar considerations apply when the mass differences have
``inverted'' hierarchy, eq.~(\ref{defih}) with $|U_{\rm e1}^2|$
playing the role of $|U_{\rm e3}^2|.$ Notice in particular that with
approximate mass degeneracy the role of the sub-dominant mixing is
almost the same for ``normal'' and ``'inverted'' hierarchy; this
should be contrasted with the conclusions for the cases [${\cal N}$]
and [${\cal I}$], when $m_1\ll (\Delta m^2_{atm})^{1/2}$.

 In the particular case of SMA solution, eq.~(\ref{smanorm}) is still
valid, with $m_1\approx m_3$ (and $U_{\rm e3}\to U_{\rm e1}$ for
``inverted'' hierarchy); hence, up to sub-dominant mixing terms ${\cal
M}_{{\rm ee}}\approx {\cal M}_{{\rm ee}}^{max} \approx m_1,$ and a
complete cancellation is impossible.

\pagebreak

\FIGURE[p]{%
\epsfig{file=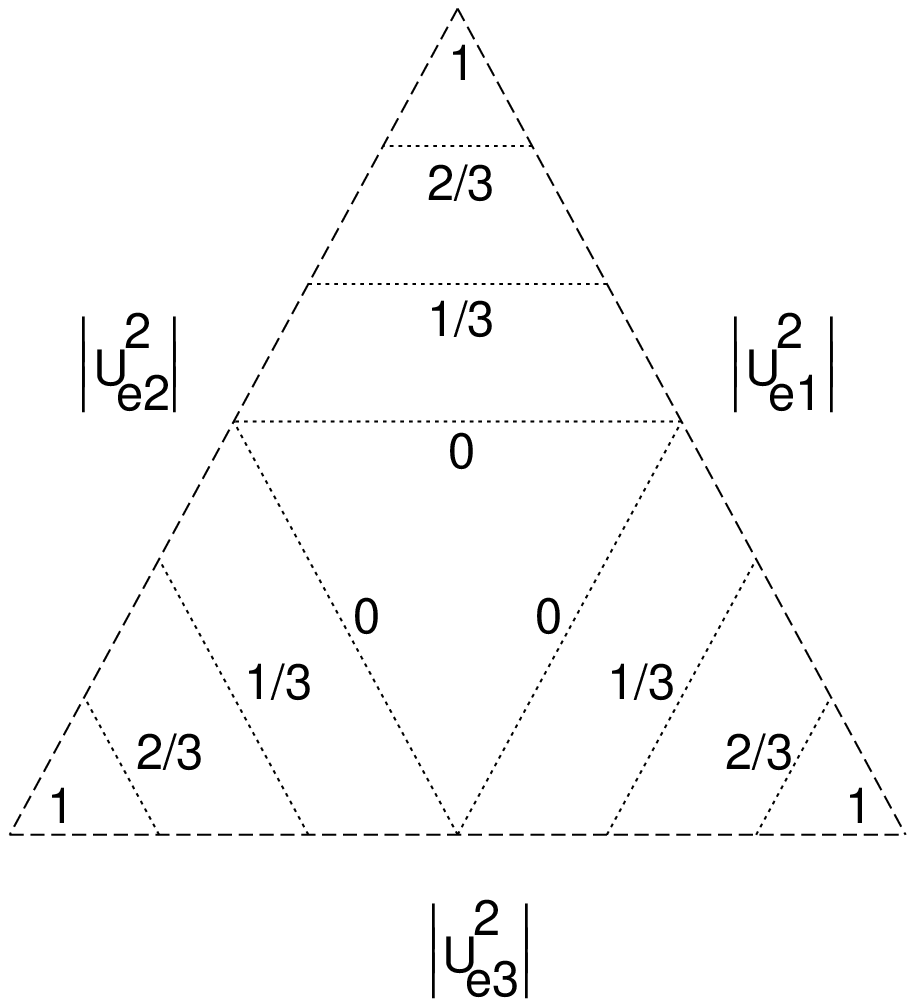, angle=270,width=17em}%
\epsfig{file=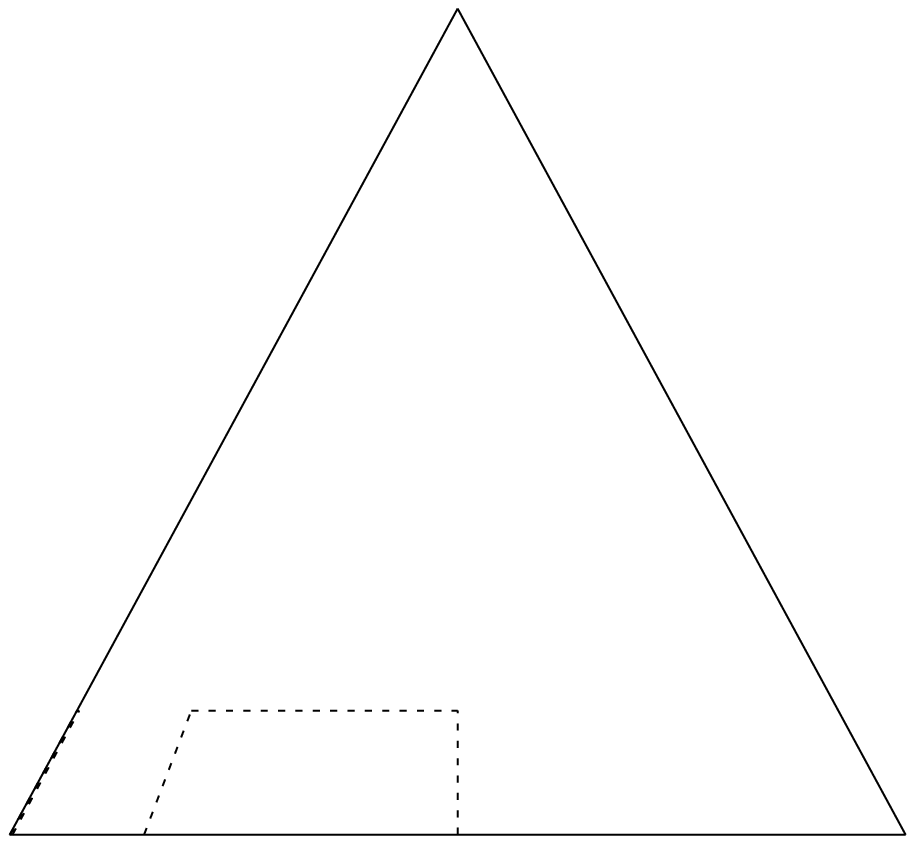, angle=270,width=17em}
\epsfig{file=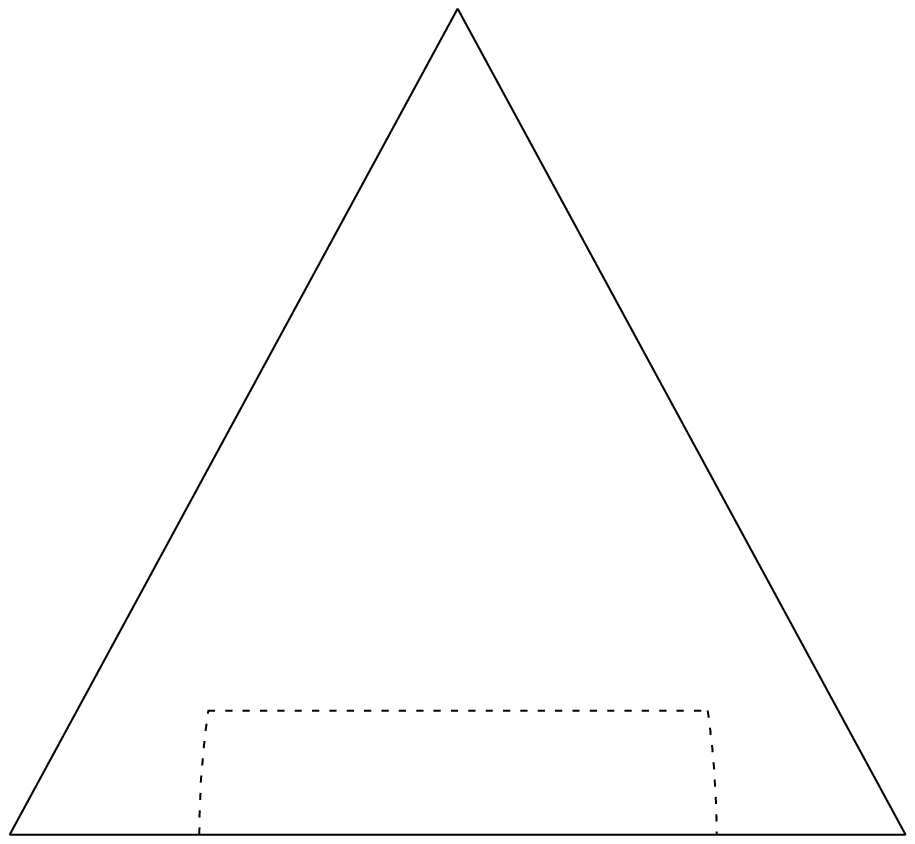, angle=270,width=17em}%
\epsfig{file=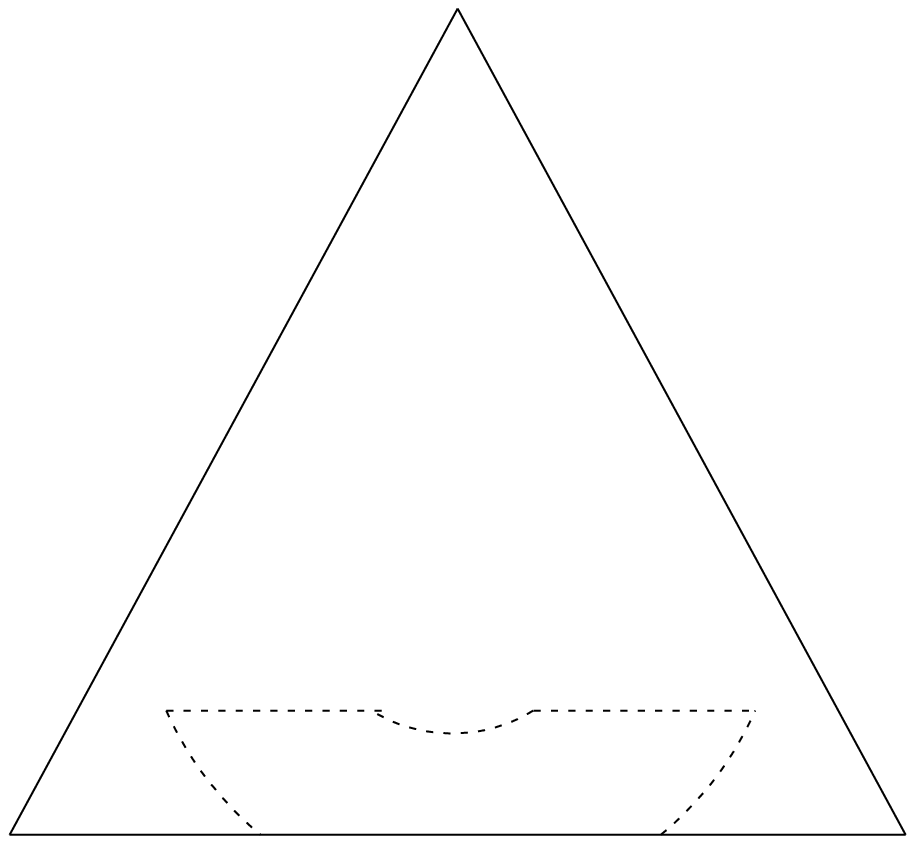,angle=270,width=17em}%
\caption{Case of ``nearly degenerate'' neutrino spectrum ({$m_1\approx
m_2\approx m_3$}). From up to down, left to right: {$1^{st}$} plot,
the minimum value of {${\cal M}_{{\rm ee}}/m_1$} represented in the
unitarity triangle.  Reference numerical values of 1, 2/3, 1/3 and 0
are indicated.  {$2^{nd}$} plot, indicative allowed regions for MSW
enhanced transitions; the SMA solution is almost superimposed to the
left--second--side, where {$U_{{\rm e}2}=0$}; {$3^{rd},$} allowed
region for vacuum oscillations; {$4^{th},$} allowed region for
averaged oscillations.  We assume {$|U_{\rm e3}^2|<0.15$}
\cite{subdmix} and ``normal'' hierarchy.  ``Inverted'' hierarchy
corresponds approximatively to a {$120^\circ$} rotation of the last
three plots.}\label{f:3}}

\clearpage

\subsection{A complementary representation}\label{sec-cr}

In order to recapitulate and confirm the results obtained in this
section, we present a complementary graphical representation.
Supposing that the mixing elements are known with good precision, we
can plot the range of values of ${\cal M}_{{\rm ee}}$ as a function of
the only residual parameter: The mass of the lightest
neutrino\footnote{In practice, this representation will be useful when
the parameters of oscillation will be known reliably.}.  This is done
in fig.~\ref{f:4}, where we assume the mass splittings $\Delta
m^2_{atm}=2\times 10^{-3}$ eV$^2$ and $\Delta m^2_{\odot}=10^{-4}$
eV$^2$ for ``normal'' and ``inverted'' hierarchy.  The mixing $|U_{\rm
e3}^2|$ (resp.\ $|U_{\rm e1}^2|$) with the heaviest (resp.\ lightest)
state is $0,2,4$ and 6 $ \times 10^{-2}$ in the 4 types of curves,
going from inner to outer ones.  We fixed $|U_{{\rm e}2}^2|=0.4$
(resp.\ $|U_{{\rm e}3}^2|=0.4$), which corresponds roughly to an LMA
solution.  The figure confirms the conclusions obtained in
section~\ref{sec-n} for the case [${\cal N}$], about the importance of
$\Delta m^2_{\odot},$ and of the small mixing element $|U_{\rm
e3}^2|.$ For the case [${\cal I}$], instead, $|U_{\rm e1}^2|$ and
$\Delta m^2_{\odot}$ are less important in agreement with the
discussion in section~\ref{sec-i}.

This representation emphasizes that also a {\em null} experimental
result may be a very important information on the massive neutrino
parameters: In fact, ${\cal M}_{{\rm ee}}^{min} \raisebox{-.4ex}
{\rlap{$\sim$}} \raisebox{.4ex}{$<$} 10^{-2}$ eV could rule out the
assumption of ``inverted'' hierarchy, see the second plot of
fig.~\ref{f:4}; or, a bound on ${\cal M}_{{\rm ee}}^{min} $ at the
$10^{-3}$ level could amount to a measurement of the lightest neutrino
mass, see the first plot of the same figure.  Unfortunately, the value
of $m_1$ determined in this way depends strongly on the parameters of
oscillation, since:
\begin{equation}
{\cal M}_{{\rm ee}}^{min}=
\left|\ |U_{{\rm e}2}^2|\ (\Delta m^2_\odot)^{1/2}
-|U_{{\rm e}3}^2| \ (\Delta m^2_{atm})^{1/2}\
\right|\ \ \ \ \mbox{for }m_1=0;
\end{equation}
so that, even in the LMA case we are considering, it will be a real
challenge to prove that $m_1\neq 0.$

\section{Concluding remarks}

\subsection{On the case 
${\cal M}_{{\rm ee}}\approx 0$}

We regarded ${\cal M}_{{\rm ee}}$ as a function of several parameters:
the mixing elements, the squared mass splittings, the mass of the
lightest neutrino and the complex phases.  Following this approach,
one may be led to wonder whether the cases when the rate is small as a
consequence of cancellations among the various parameters are (in some
sense) ``natural''.

We show here how the smallness can arise in a ``natural'' manner.  Let
us postulate that the neutrino mass matrix has a hierarchical
structure, analogous to the structure of the Yukawa couplings of the
charged fermions.

\pagebreak

\FIGURE[p]{%
\epsfig{file=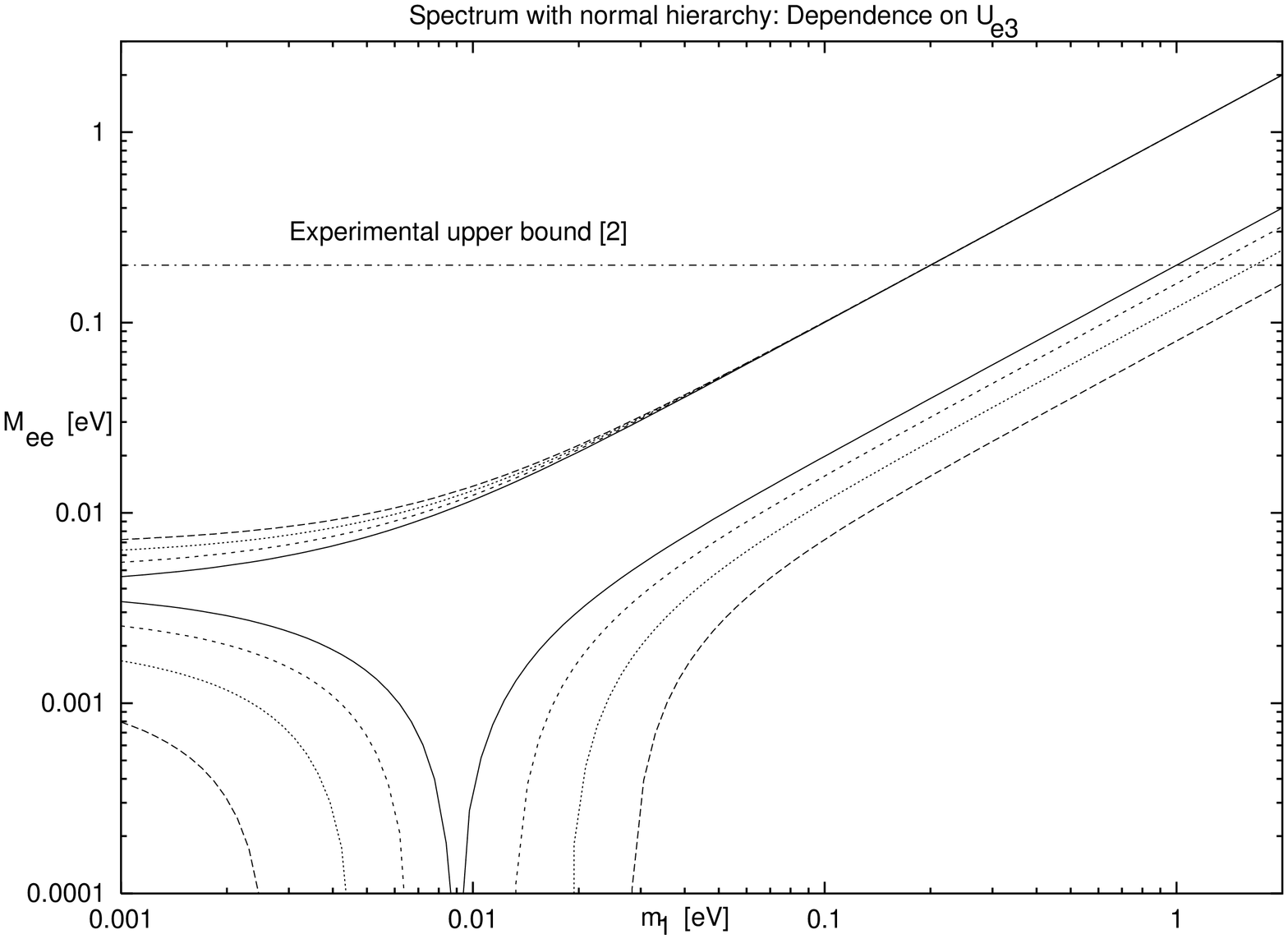, angle=270,width=33em}
\epsfig{file=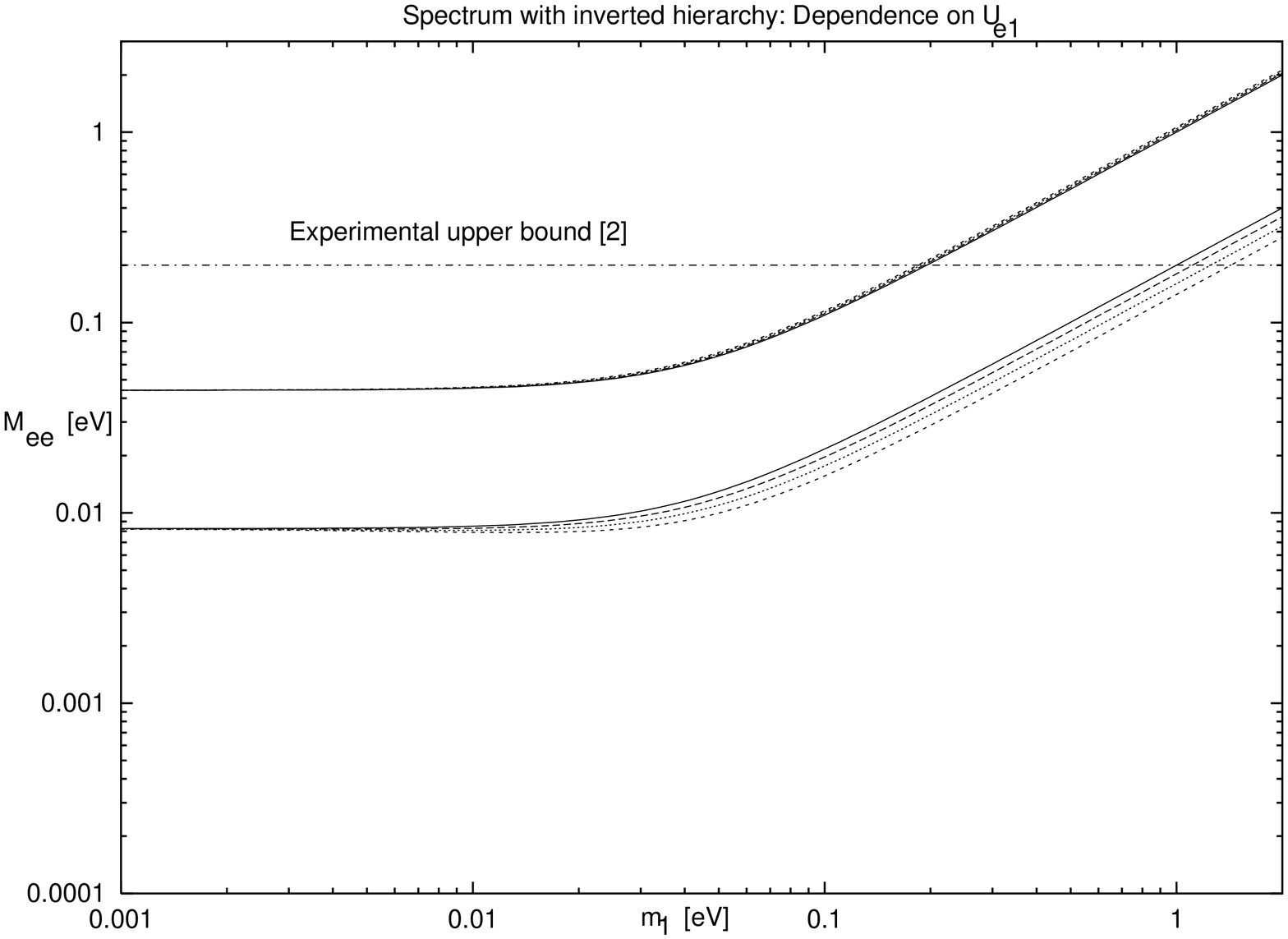, angle=270,width=33em}%
\caption{Range of values of {${\cal M}_{{\rm ee}}.$} The four upper
curves in each plot represent {${\cal M}_{{\rm ee}}^{max},$} the four
lower ones {${\cal M}_{{\rm ee}}^{min}$} --- eqs.\ ({\ref{eq2}}),
({\ref{eq3}}).  The four types of curve differ for the values of the
sub-dominant mixing element: For ``normal'' hierarchy, {$|U_{\rm
e3}^2| =0,2,4$} and {$6 \times 10^{-2},$} going from inner curves
(continuous line) to outer ones (long-dashed); same values, but for
{$|U_{\rm e1}^2|,$} in the case of ``inverted''
hierarchy.\label{f:4}}}

\clearpage

\noindent In this case, we can expect that the ``ee-entry'' of the
neutrino mass matrix ($={\cal M}_{{\rm ee}}$) is the smallest one, and
also ${\cal M}_{{\rm ee}}\ll (\Delta m^2_{atm})^{1/2}.$ This is what
happens in the two models of references~\cite{yanaram}, where:
\begin{equation}
{\cal M}_{{\rm ee}}\approx 
(\Delta m^2_{atm})^{1/2}\times 
(\sin\theta_C)^{2 n} ;
\end{equation}
$\theta_C$ is the Cabibbo angle, and $n=2,3$ in the two models
respectively.  The value of ${\cal M}_{{\rm ee}}$ in these models is
rather small (see also~\cite{yb}).  Although the contribution from
third family is modest, LMA solutions with relatively large mass
splittings are possible in this type of models~\cite{large}, which
{\em a priori} may imply much larger values of ${\cal M}_{{\rm ee}},$
as remarked for the case of section~\ref{sec-n}.  Thus, these models
provide examples of cases when ${\cal M}_{{\rm ee}}$ is small as a
consequence of cancellations among the various contributions.

In another sense, the statement ${\cal M}_{{\rm ee}}\approx 0$ is
surely ``natural'' in a standard model framework, since at one loop
level the radiative corrections are tiny: $\sim y_e^2/(4\pi)^2\sim
5\times 10^{-14},$ where $y_e$ is the electron Yukawa coupling.

\subsection{What is the maximum value of ${\cal M}_{{\rm ee}}?$}

Let us briefly summarize the results of section~\ref{secdef}, about an
aspect of importance for experimental search: The maximum value of
${\cal M}_{{\rm ee}}$ that we can {\em a priori} expect.

For given mixing elements, ${\cal M}_{{\rm ee}}^{max}$ {\em increases}
passing from the cases discussed in sections~\ref{sec-n} (case [${\cal
N}$]) to section~\ref{sec-i} (case [${\cal I}$]), and finally to
section~\ref{sec-d} (case [${\cal D}$]).  Indeed, ${\cal M}_{{\rm
ee}}^{max}$ reaches at most the $10^{-2}$ eV level in case [${\cal
N}$], depending on the sub-dominant mixing element $|U_{{\rm e}3}^2|$
and on the scenario of oscillation (eq.~(\ref{normsig})); it can be of
the order of 3 to $9\times 10^{-2}$ eV in the case [${\cal I}$],
depending on the size of $\Delta m^2_{atm}$ (eq.~(\ref{invsig}));
finally, ${\cal M}_{{\rm ee}}^{max}$ can be as large as the
experimental upper limit of 0.2 eV in the case [${\cal D}$].  In this
sense, the {\em a priori} hope of a positive experimental result
increases when going from [${\cal N}$] to [${\cal I}$], and from
[${\cal I}$] to [${\cal D}$]\footnote{On the contrary, one might argue
that the case [${\cal N}$] is more likely than [${\cal I}$], and this
latter more likely than [${\cal D}$], again on the basis of an analogy
between the neutrino spectrum and the spectra of the charged
fermions.}.

\subsection{Studies of neutrino oscillations and search for 
$0\nu 2\beta$ decay}

We have shown that the parameters of oscillations are strictly related
to the possible value of the $0\nu2\beta$ decay rate.  However, the
dependence on the type of spectrum is also essential.  We summarize
here some results of special interest (making reference for details to
the previous section):

\noindent $\bullet$ For the small angle MSW solution, ${\cal M}_{\rm
ee}$ is quite large for ``inverted'' hierarchy in the case $m_1\ll
(\Delta m^2_{atm} )^{1/2},$ see eq.~(\ref{smainverted}); for
``normal'' hierarchy, we have instead eq.~(\ref{smanorm}), which is
smaller than the previous case by a factor of $|U_{\rm e3}^2|$ when
$m_1$ is small, and possibly even smaller (eq.~(\ref{smanorm})).\\
$\bullet$ For the large mixing angle MSW solution, contributions from
``solar'' frequency, order $(\Delta m^2_{\odot} )^{1/2}$ are {\em not}
negligible, and they may lead to cancellations (or enhancements)
depending on the size of $|U_{\rm e3}^2|$ in the case of ``normal''
hierarchy (sections~\ref{sec-n} and \ref{sec-cr}).\\ 
$\bullet$ For VO solution, and
``normal'' hierarchy, the dependence of ${\cal M}_{\rm ee}^{min}$ on
$|U_{\rm e3}^2|$ in quite appreciable (section\ \ref{sec-n}).\\
$\bullet$ For ``inverted'' hierarchy, cancellations are not easy to
obtain if $m_1$ is small in comparison with $(\Delta m^2_{atm}
)^{1/2},$ except for solutions of the solar neutrino problem with
almost maximal mixing angles (section~\ref{sec-i}).\\ $\bullet$
Largest values of ${\cal M}_{\rm ee}$ are taken in the case of
``nearly degenerate'' spectrum, $m_1\gg (\Delta m^2_{atm} )^{1/2}$
(section~\ref{sec-d}).  In this extreme case, cancellations are
possible especially for quite large mixing angle solutions, with
relevant dependence on the size of the sub-dominant mixing, for both
``normal'' and ``inverted'' hierarchies.

\subsection{Conclusions and perspectives}

In this work, we discussed the interplay between the studies of
neutrino oscillations and the search for $0\nu 2\beta$ decay.  We
introduced new graphical representations, aimed at clarifying the
relations between the neutrino spectra, the scenarios of oscillations
and the rate of the neutrinoless double beta decay.  For the
perspectives, it has to be noticed that the present information on
massive neutrinos is compatible with quite different oscillations
scenarios and neutrino spectra.  Future experiments aiming at a signal
of the $0\nu 2\beta$ process above the $10^{-2}$ eV level
\cite{genius} will have an important role in deciding among the
alternative possibilities.

\acknowledgments 

I thank R.\ Barbieri, C.\ Giunti, M.\ Maris and A.\ Yu.\ Smirnov for
useful discussions, and the Referee of the work for having suggested
important improvements.  Earlier accounts were presented in
\cite{prevacc}.

\end{document}